\documentclass[11pt]{article}

\usepackage[final]{acl}

\usepackage{times}
\usepackage{latexsym}
\usepackage{amssymb}
\usepackage{threeparttable}

\usepackage[T1]{fontenc}

\usepackage[utf8]{inputenc}
\usepackage{booktabs}
\usepackage{siunitx}
\usepackage{multirow}
\usepackage{listings}
\usepackage{amsmath}
\usepackage{bm}
\usepackage{microtype}
\usepackage{makecell}
\usepackage{inconsolata}
\usepackage[table]{xcolor}

\usepackage{graphicx}

\usepackage{enumitem}

\setlist[itemize]{leftmargin=1.5em, itemsep=1pt, topsep=1pt}
\definecolor{llm}{HTML}{6259F2}
\definecolor{ner}{HTML}{80BD29}

\title{From Text to Alpha: Can LLMs Track \\ Evolving Signals in Corporate Disclosures?}

\author{
  \textbf{Chanyeol Choi\textsuperscript{1}\thanks{Corresponding author.}},
  \textbf{Yoon Kim\textsuperscript{2}},
  \textbf{Yu Yu\textsuperscript{3}},
  \textbf{Young Cha\textsuperscript{4}},
  \textbf{V. Zach Golkhou\textsuperscript{5}},
  \textbf{Igor Halperin\textsuperscript{6}}, \\
  \textbf{Georgios Papaioannou\textsuperscript{7}},
  \textbf{Minkyu Kim\textsuperscript{8}},
  \textbf{Zhangyang Wang\textsuperscript{9}},
  \textbf{Jihoon Kwon\textsuperscript{1}},
  \textbf{Minjae Kim\textsuperscript{1}}, \\
  \textbf{Alejandro Lopez-Lira\textsuperscript{10}},
  \textbf{Yongjae Lee\textsuperscript{11}}
  \\
  \\
  \textsuperscript{1}LinqAlpha \quad
  \textsuperscript{2}MIT \quad
  \textsuperscript{3}BlackRock \quad
  \textsuperscript{4}Blackstone \quad
  \textsuperscript{5}J.P. Morgan \\
  \textsuperscript{6}Fidelity Investments \quad
  \textsuperscript{7}Qube Research \& Technologies \quad
  \textsuperscript{8}State Street Investment Management \\
  \textsuperscript{9}University of Texas at Austin \quad
  \textsuperscript{10}University of Florida \quad
  \textsuperscript{11}UNIST
}

\begin{document}
\maketitle
\begin{abstract}
Natural language processing (NLP) has been widely used in quantitative finance, but traditional methods often struggle to capture rich narratives in corporate disclosures, leaving potentially informative signals under-explored.
Large language models (LLMs) offer a promising alternative due to their ability to extract nuanced semantics. In this paper, we ask whether semantic signals extracted by LLMs from corporate disclosures predict alpha, defined as abnormal returns beyond broad market movements and common risk factors. We introduce a simple framework, \emph{LLM as extractor, embedding as ruler}, which extracts context-aware, metric-focused textual spans and quantifies semantic changes across consecutive disclosure periods using embedding-based similarity.
This allows us to measure the degree of metric shifting -- how much firms move away from previously emphasized metrics, referred as moving targets.
In experiments with portfolio and cross-sectional regression tests against a recent NER-based baseline, our method achieves more than twice the risk-adjusted alpha and shows significantly stronger predictive power.
Qualitative analysis suggests that these gains stem from preserving contextual qualifiers and filtering out non-metric terms that keyword-based approaches often miss.
\end{abstract}

\section{Introduction}

Natural language processing (NLP) has been widely adopted for quantitative investment in finance due to its ability to identify complex patterns and relationships in financial text~\citep{li2023large, lee2024survey, cao2025deep}.
Traditional approaches based on named entity recognition (NER) and text classification have been shown to extract alpha from financial text~\citep{frankel1999empirical, tetlock2007giving, matsumoto2011makes, loughran2011liability, loughran2016textual}.
In this context, alpha refers to abnormal returns that are not explained by broad market movements or common risk factors.
However, these methods often struggle to capture rich semantics and nuanced sentiment in corporate disclosures, leaving potentially informative signals under-explored.
Large language models (LLMs) have emerged as a promising alternative because they can extract subtle meaning and contextual qualifiers that traditional pipelines tend to miss~\citep{chen2022expected, lopez2023can, guo2024fine}.

In this paper, we ask a concrete question: do semantic signals extracted by LLMs from corporate disclosures predict alpha?
We address this question by quantifying semantic changes across disclosures over time, focusing on how firms shift the performance metrics they emphasize across consecutive disclosure periods.
To this end, we introduce a simple and general framework, \emph{LLM as extractor, embedding as ruler}.
The two components address complementary challenges.
First, we use an LLM to extract metrics from each disclosure, preserving contextual qualifiers that NER-based approaches often collapse.
Second, we compare extracted metrics across periods using embedding-based semantic similarity, which can identify equivalent metrics expressed differently.
This similarity serves as a ruler for measuring how the underlying metrics evolve over time.

To validate our framework, we compare against the recent baseline proposed by \citet{cohen2024moving}.
That work introduces an NER-based metric called \emph{moving targets}, which captures how firms strategically shift the performance metrics they highlight across disclosures.
Such metric shifting has been shown to contain predictive information about future stock returns.
Using an LLM-based analogue of moving targets derived from our framework, we conduct experiments in a realistic setting and show that our approach captures information that traditional NER-based methods fail to quantify, leading to improved prediction of excess returns.

Our empirical evaluation follows two standard protocols in empirical finance.
First, we construct portfolios that buy shares with low metric shifting and sell shares with high metric shifting, motivated by prior evidence that greater metric shifting predicts lower future returns.
We then measure portfolio performance after controlling for common market factors~\citep{fama2015five}.
Our method achieves an alpha more than twice as high as that of the baseline.
Second, we evaluate return predictability using cross-sectional regressions of next-month stock returns~\citep{fama1973risk}.
Across these tests, our approach exhibits stronger predictive power than the baseline.
In addition, qualitative analysis indicates that these improvements stem from capturing semantic equivalence and contextual qualifiers that keyword-based approaches often overlook.
Together, these results demonstrate that LLMs provide a robust mechanism for tracking evolving signals in corporate disclosures and for predicting abnormal returns.

\section{Related Work}
\paragraph{Financial LLMs.}
Recently, domain-adapted pretrained language models and LLMs have reported performance gains on standard financial NLP tasks such as sentiment classification and NER~\citep{kaur2023refind, du2024financial}.
Early work focused on further pre-training or fine-tuning BERT-based models on financial text, achieving strong results on benchmarks such as sentiment classification~\citep{araci2019finbert, liu2021finbert}.
Subsequent efforts expanded to large-scale finance-specific LLMs, enabling evaluation across a broader range of financial NLP tasks beyond traditional sentiment and NER, including question answering~\citep{liu2023fingpt, wu2023bloomberggpt, islam2023financebench}.
A parallel line of work explores instruction-tuning approaches evaluating multi-task and zero-shot generalization~\citep{xie2023pixiu, zhang2023instruct}.

\paragraph{Text Analysis on Financial Text.}
A large body of research in finance has shown that financial text contains information predictive of future stock returns~\citep{tetlock2007giving, loughran2011liability, loughran2016textual}.
In particular, earnings calls have received attention as a primary channel through which managers communicate firm performance~\citep{frankel1999empirical, matsumoto2011makes}.
Studies analyzing individual call transcripts have shown that managerial language -- such as tone or readability -- can predicts excess returns~\citep{price2012earnings, mayew2012power, fu2021information}.
More recent work shifts focus from analyzing individual documents to examining changes in content across earnings call transcripts~\citep{cohen2024moving}, showing that textual modifications over time carry incremental information for return predictability.

In parallel, a growing literature explores the use of LLMs in finance, applying LLMs to news articles, annual reports, and earnings call transcript to improve return prediction and portfolio performance~\citep{chen2022expected, lopez2023can, guo2024fine, magner2025decoding}.
However, existing studies have primarily validated LLMs on single-document tasks; their effectiveness in tracking semantic changes across documents over time remains unexplored.

\section{Backgrounds}

\begin{figure}[t]
    \centering
    \includegraphics[width=0.4\textwidth]{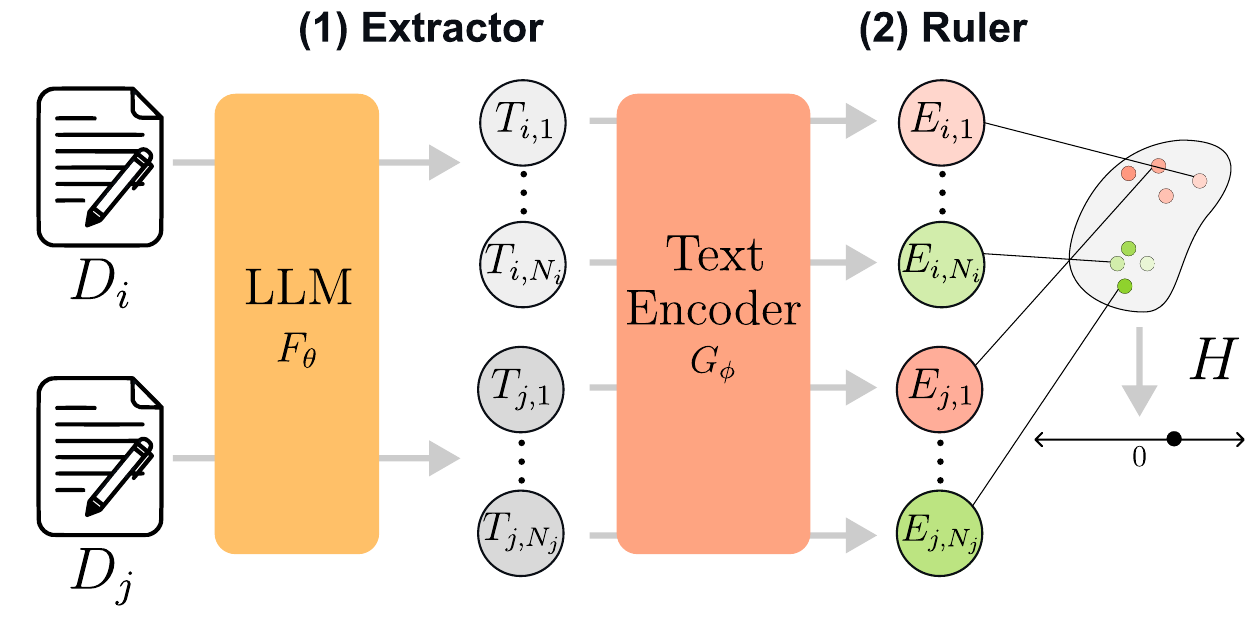}
    \caption{
Overview of our framework, \emph{LLM as extractor, embedding as ruler}.
Given two documents $D_i$ and $D_j$ from different time periods, an LLM $F_\theta$ first acts as an \emph{extractor}, identifying context-aware contents from each document.
A pretrained text encoder $G_\phi$ then serves as a \emph{ruler}, mapping extracted contents into a shared embedding space and measuring their semantic similarity to quantify how extracted contents persistence, disappearance, and emergence.
}
    \label{fig:overview}
\end{figure}

\paragraph{Earnings Calls and Performance Metrics.}
Publicly traded companies hold quarterly earnings conference calls to discuss financial results with analysts and investors~\citep{tasker1998bridging}.
These calls typically consist of two parts: a presentation by management summarizing the quarter's performance, followed by a question-and-answer session with analysts~\citep{frankel1999empirical}.
During these calls, managers highlight specific metrics -- quantitative measures such as revenue or product-specific figures -- to frame the firm's progress.
Because these metrics are chosen by management rather than externally mandated, managers have discretion over which metrics to emphasize and how prominently to feature them~\citep{price2012earnings}.

\begin{figure*}[t]
    \centering
    \includegraphics[width=0.8\textwidth]{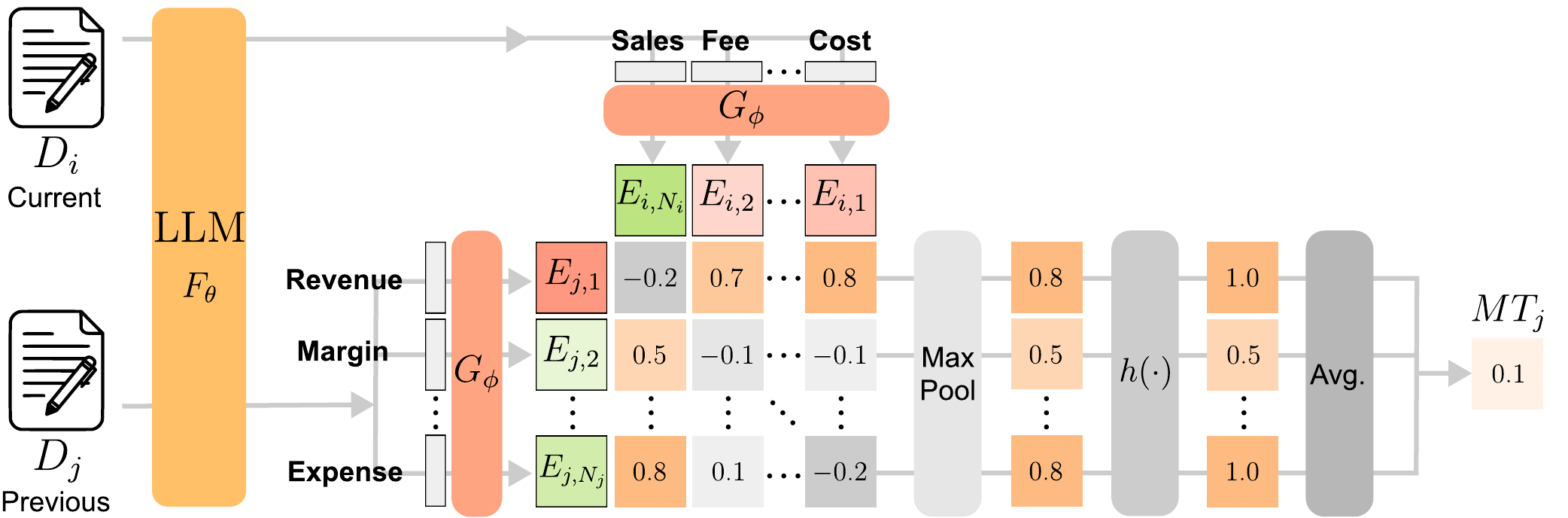}
    \caption{
Illustration of the scoring pipeline using embedding similarity.
Suppose that two sets of metrics have already been extracted from two transcripts $D_i$ (current) and $D_j$ (previous).
First, we encode each metric into a shared embedding space using a pre-trained text encoder $G_\phi$, obtaining embeddings $\bm{E}_{i,n_i}$ and $\bm{E}_{j,n_j}$ for quarter indices $i$ and $j$.
For each metric $\bm{E}_{j,n_j}$, we then compute its maximum cosine similarity $\hat{S}_{n_j}$ to any metric at quarter $i$, which tests whether a semantically corresponding metric is still present.
Next, we apply thresholding $h(\cdot)$ to suppress weak matches and treat only sufficiently similar pairs as retained.
Finally, we aggregate these similarities using an averaging operator, quantifying the fraction of previously discussed metrics that are no longer emphasized.
}
    \label{fig:case}
\end{figure*}

\paragraph{Moving Targets.}
\citet{cohen2024moving} document that managers strategically shift their emphasis on metrics over time.
When a previously highlighted metric becomes difficult to sustain -- for example, when sales growth stalls -- managers often pivot to emphasizing different metrics, such as cost savings or strategic investments.
This phenomenon, termed \emph{moving targets}, is formally measured as the fraction of previously discussed metrics that are absent in the current period.
In their approach, metrics are extracted using spaCy~\citep{honnibal2017spacy} NER and compared across periods via rule-based string matching; we refer to this as the \textcolor{ner}{\emph{NER-based}} method.

Following their definition, we refer to this measure as the moving targets (MT) score, computed as
\begin{equation}
\text{MT}_i = \frac{\sum (\text{Missing Metrics}_i \mid \text{Metrics}_{i-4})}{\sum \text{Metrics}_{i-4}}
\end{equation}
where $i$ indexes the fiscal quarter, $\text{Metrics}_i$ is the set of performance metrics mentioned in quarter $i$, and $i-4$ refers to the same fiscal quarter one year prior (e.g., Q1 2024 vs.\ Q1 2023).
$\text{Missing Metrics}_i$ denotes metrics that were mentioned in $i-4$ but are absent in $i$.
Higher values indicate greater departure from previously emphasized metrics.

\section{Method}

To test whether LLMs can better capture the moving targets, we employ the \emph{LLM as extractor, embedding as ruler} framework.
Fig.~\ref{fig:overview} shows the overall pipeline.
The LLM serves as an \emph{extractor} that identifies metrics from transcripts while preserving contextual qualifiers.
The embedding model serves as a \emph{ruler} that measures semantic distance between metrics across periods, enabling comparison even when equivalent metrics are expressed differently.

\paragraph{Extractor: Metric Extraction.}
Let $D_{i}$ denote the earnings call transcript of a firm at quarter index $i$.
We extract performance metrics using a pre-trained LLM $F_{\theta}$: given a task-specific prompt and the entire transcript $D_{i}$ as input, the model outputs a set of mentioned metrics.
We denote the obtained set as $\mathcal{T}_{i} = \{T_{i,1}, \dots, T_{i,N_i}\}$, where $N_i$ denotes the number of metrics extracted from quarter $i$.
Unlike NER-based approaches that often extract only generic terms, the LLM can capture metrics expressed through complex phrases while preserving contextual qualifiers (e.g., ``North America cloud revenue'' rather than just ``revenue'').

\paragraph{Ruler: Semantic Similarity Scoring.}
The original moving targets score counts the fraction of metrics from $j=i{-}4$ that are absent at $i$ via keyword matching.
However, this fails when semantically equivalent metrics are expressed differently across quarters (e.g., ``sales growth'' vs.\ ``revenue increase'').
To address this, we use embeddings as a ruler to measure semantic distance.
We illustrate the concept of our framework in Fig.~\ref{fig:case}.

We encode each extracted metric using a pretrained text encoder $G_\phi$, obtaining embeddings $\bm{E}_{i,n_i} = G_\phi(T_{i,n_i}) \in \mathbb{R}^d$.
For each previous metric $T_{j,n_j}$ and its embedding $\bm{E}_{j,n_j}$, we compute its maximum cosine similarity to any current metric:
\begin{equation}
\hat{S}_{n_j} = \max_{n_i} \cos(\bm{E}_{i,n_i}, \bm{E}_{j,n_j})
\end{equation}

Note that raw cosine similarity does not directly correspond to semantic equivalence: pairs below a certain threshold are effectively dissimilar, while pairs above another threshold are highly similar.
To reflect this, we apply a piecewise-linear transformation $h(\cdot)$:
\begin{equation}
\label{eq:threshold}
{S}_{n_j} = h(\hat{S}_{n_j}) = \begin{cases}
0 & \text{if } \hat{S}_{n_j} \leq \alpha \\
\frac{\hat{S}_{n_j} - \alpha}{\beta - \alpha} & \text{if } \alpha < \hat{S}_{n_j} < \beta \\
1 & \text{if } \hat{S}_{n_j} \geq \beta
\end{cases}
\end{equation}
This maps low similarities to 0 (dropped) and high similarities to 1 (retained), reducing noise from ambiguous middle-range scores.

Finally, the embedding-based MT score is defined as:
\begin{equation}
MT_{i} = 1 - \frac{1}{N_j} \sum_{n_j=1}^{N_j} {S}_{n_j}
\end{equation}
Since ${S}_{n_j} \approx 1$ when a metric is retained and ${S}_{n_j} \approx 0$ when dropped, this formulation yields the fraction of metrics that are absent -- matching the original definition.
We refer to this approach as the \textcolor{llm}{\emph{LLM-based}} method.

\section{Experiments}

In this section, we empirically evaluate whether LLM-based tracking of moving targets leads to better prediction of excess returns.
Sec.~\ref{subsec:exp_setup} describes our experimental setup.
Sec.~\ref{subsec:results} compares predictive performance quantitatively and examines the sources of improvement qualitatively.
Codes are available at \href{https://anonymous.4open.science/r/Evolving-Signals-in-Corporate-Disclosures-8775/README.md}{this GitHub repository}.

\subsection{Experimental Setup}
\label{subsec:exp_setup}
\paragraph{Data.}
Our experimental setup follows~\citet{cohen2024moving}.
Our sample covers firms listed in the S\&P 100 index from January 2010 to December 2024, yielding 5,615 firm-quarter observations across 64 quarters.
While the original study uses a broader universe, we focus on S\&P 100 firms due to computational cost constraints associated with LLM-based extraction.
This sample remains sufficiently large to detect economically meaningful effects and covers a diverse set of industries.

\paragraph{Evaluation Protocols.}
To assess whether the moving targets scores carry economically meaningful predictive power, we adopt the evaluation framework of \citet{cohen2024moving}, which consists of two complementary analyses.

First, we construct a portfolio strategy based on the moving targets score and compare portfolio returns.
Following~\citet{cohen2024moving}, we sort firms into quintiles based on their most recent moving targets score and form equal-weighted portfolios.
Q1 contains firms with the lowest scores, while Q5 contains firms with the highest scores.
We then compute the return spread between Q5 and Q1.
A significantly negative Q5--Q1 spread indicates that firms dropping more targets subsequently underperform.
To account for other factors that may drive returns, we report risk-adjusted alphas using the Fama--French three-factor and five-factor models~\citep{fama1993common, fama2015five}.

Second, we test whether the moving targets score significantly predicts next-month stock returns using Fama-MacBeth cross-sectional regressions~\citep{fama1973risk}, a standard protocol in empirical finance for testing whether a variable predicts future returns.
Each month, we regress individual stock returns on the moving targets score and control variables.
We then average the monthly coefficient estimates across time.
A significantly negative coefficient on the moving targets score indicates that more target shifting predicts lower subsequent returns.

\begin{table}[t]
\centering
\caption{Statistics of dataset. We use earnings call transcripts from firms listed in S\&P 100 index, spanning January 2010 to December 2024.}
\resizebox{0.9\columnwidth}{!}
{
\label{tab:data}
\begin{tabular}{lr}
\toprule
\textbf{Statistic} & \textbf{Value} \\
\midrule
Sample period & Jan. 2010 -- Dec. 2024 \\
Number of quarters & 64 \\
Number of firms & 100 \\
Firm-quarter observations & 5,615 \\
Returns observations & 16,675 \\
\bottomrule
\end{tabular}
}
\end{table}
\begin{table}[t]
\centering
\caption{
Comparison of the top 10 most frequent missing targets extracted by the \emph{NER-based} and \emph{LLM-based} methods.
}
\Huge
\label{tab:freq}
\resizebox{1.0\columnwidth}{!}{
\begin{tabular}{r l r | r l r}
\toprule
\multicolumn{3}{c}{\textcolor{ner}{\textbf{NER-based Method}}} & \multicolumn{3}{c}{\textcolor{llm}{\textbf{LLM-based Method}}} \\
\cmidrule(lr){1-3} \cmidrule(lr){4-6}
\textbf{Rank} & \textbf{Target} & \textbf{Frequency}
& \textbf{Rank} & \textbf{Target} & \textbf{Frequency} \\
\midrule
1  & the \%               & 4460 & 1  & market share             & 706 \\
2  & a \% increase        & 3845 & 2  & share repurchases        & 589 \\
3  & the range            & 3474 & 3  & revenue growth           & 582 \\
4  & \% growth            & 3456 & 4  & pricing                  & 562 \\
5  & q3                   & 3006 & 5  & cash flow                & 535 \\
6  & q4                   & 2853 & 6  & quarterly revenue growth & 517 \\
7  & the \% range         & 2801 & 7  & yearly guidance          & 484 \\
8  & year                 & 2716 & 8  & free cash flow           & 480 \\
9  & the quarter          & 2499 & 9  & dividend                 & 458 \\
10 & share                & 2400 & 10 & quarterly earnings per share & 455 \\
\bottomrule
\end{tabular}
}
\end{table}

\begin{figure*}[t]
    \centering
    \includegraphics[width=1.0\textwidth]{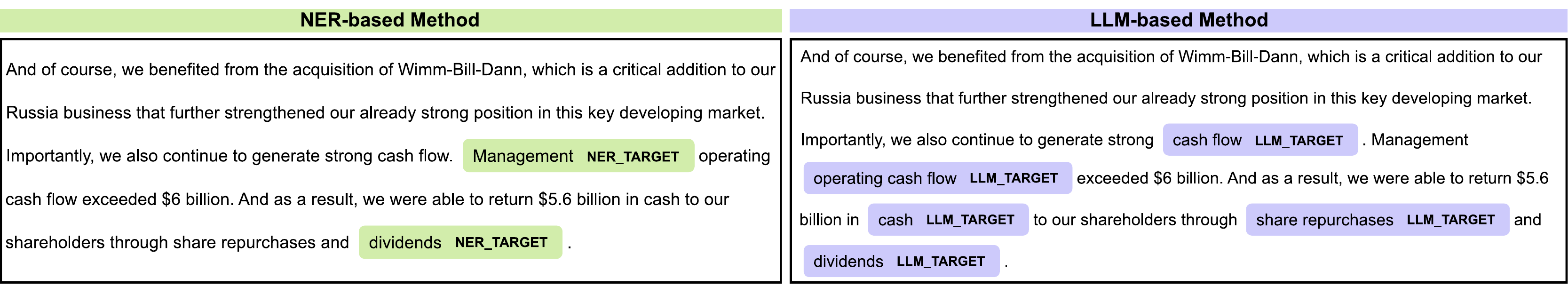}
    \caption{
Comparison of extracted metrics between NER-based and LLM-based methods on a PepsiCo Q4 2011 earnings call transcript.
The panel shows the same transcript excerpt with NER extractions (\textbf{left}) and LLM extractions (\textbf{right}).
\textbf{Green} highlights indicate NER-extracted terms; \textbf{purple} highlights indicate LLM-extracted metrics.
Given the same passage discussing cash flow performance, NER extracts only ``Management'' and ``dividends'', missing the key financial metrics mentioned.
In contrast, LLM correctly identifies ``cash flow'', ``operating cash flow'', ``cash'', ``share repurchases'', and ``dividends''.
}
    \label{fig:qual}
\end{figure*}

\begin{table*}[t]
\centering
\caption{
Calendar-time portfolio excess returns and 3-factor / 5-factor alphas based on \emph{Moving Targets}.
(A) reports results using the NER-based extraction method, while (B) reports results using the LLM-based extraction method.
$t$-statistics are in parentheses. $^{*}$, $^{**}$, and $^{***}$ indicate significance at the 10\%, 5\%, and 1\% levels, respectively.
}
\label{tab:portfolio}
\tiny
\renewcommand{\arraystretch}{0.3}
\resizebox{0.85\textwidth}{!}{%
\begin{tabular}{lrrrrrr}
\toprule
 & \textbf{Q1} & \textbf{Q2} & \textbf{Q3} & \textbf{Q4} & \textbf{Q5} & \textbf{Q5--Q1} \\
\midrule
\multicolumn{7}{l}{(A) \textcolor{ner}{\emph{NER-based method}}} \\

Excess Return
 & 0.0156$^{***}$ & 0.0131$^{***}$ & 0.0111$^{***}$ & 0.0132$^{***}$ & 0.0135$^{***}$ & -0.0031 \\
 & (4.38) & (3.85) & (3.19) & (3.99) & (3.94) & (-1.48) \\

\cmidrule(lr){2-7}
3-Factor Alpha
 & 0.0047$^{***}$ & 0.0027$^{**}$ & 0.0017 & 0.0040$^{***}$ & 0.0039$^{**}$ & -0.0018 \\
 & (3.43) & (2.12) & (1.15) & (2.88) & (2.47) & (-0.89) \\

\cmidrule(lr){2-7}
5-Factor Alpha
 & 0.0045$^{***}$ & 0.0028$^{**}$ & 0.0016 & 0.0040$^{***}$ & 0.0041$^{**}$ & -0.0014 \\
 & (3.26) & (2.19) & (1.07) & (2.79) & (2.54) & (-0.70) \\

\midrule
\multicolumn{7}{l}{(B) \textcolor{llm}{\emph{LLM-based method}}} \\

\rowcolor{gray!20}
Excess Return
 & 0.0148$^{***}$ & 0.0131$^{***}$ & 0.0128$^{***}$ & 0.0132$^{***}$ & 0.0117$^{***}$ & \textbf{-0.0041$^{**}$} \\
\rowcolor{gray!20}
 & (4.75) & (3.84) & (3.60) & (3.65) & (3.50) & (-2.08) \\

\cmidrule(lr){2-7}
\rowcolor{gray!20}
3-Factor Alpha
 & 0.0056$^{***}$ & 0.0036$^{***}$ & 0.0024$^{*}$ & 0.0029$^{*}$ & 0.0018 & \textbf{-0.0048$^{**}$} \\
\rowcolor{gray!20}
 & (3.89) & (2.63) & (1.89) & (1.91) & (1.22) & (-2.40) \\

\cmidrule(lr){2-7}
\rowcolor{gray!20}
5-Factor Alpha
 & 0.0055$^{***}$ & 0.0038$^{***}$ & 0.0026$^{**}$ & 0.0032$^{**}$ & 0.0013 & \textbf{-0.0052$^{**}$} \\
\rowcolor{gray!20}
 & (3.71) & (2.78) & (2.02) & (2.02) & (0.87) & (-2.55) \\

\bottomrule
\end{tabular}%
}
\end{table*}

\paragraph{Implementation Details.}
For metric extraction, we use \texttt{Gemini-2.5-Pro}, a state-of-the-art LLM with strong instruction-following capabilities.
For semantic comparison, we use \texttt{text-embedding-3-large}, a high-performing general-purpose text encoder.
Since our goal is to test whether LLM-based extraction outperforms NER-based methods---rather than to identify the optimal model---we use a single strong model for each component and leave model comparisons to future work.
We set $\alpha=0.4$ and $\beta=0.6$ for thresholding in Eq.~\ref{eq:threshold}, selected from a sweep at 0.2 intervals based on both predictive performance and interpretability of the resulting similarity boundaries.
The full extraction prompt is provided in Appendix Tab.~\ref{tab:prompts}.

\subsection{Results}
\label{subsec:results}

\paragraph{NER and LLM extract qualitatively different metrics at the aggregate level.}
Tab.~\ref{tab:freq} compares the top 10 most frequent missing targets extracted by each method.
The NER-based method predominantly captures surface-level patterns such as ``the \%'', ``a \% increase'', and ``the range'', which lack context and do not correspond to meaningful performance metrics.
In contrast, the LLM-based method extracts concrete business metrics such as ``market share'', ``revenue growth'', ``cash flow'', and ``free cash flow'' -- terms that align with the definition of forward-looking performance indicators.
This difference suggests that LLM-based extraction better leverages contextual understanding to identify targets that reflect genuine strategic emphasis by firms.

\paragraph{Individual examples confirm that LLM captures metrics NER misses.}
Fig.~\ref{fig:qual} illustrates a representative example from an earnings call transcript.
Given the same passage discussing cash flow performance, the NER-based method extracts only ``Management'' and ``dividends'', missing the key financial metrics mentioned.
The LLM-based method correctly identifies ``cash flow'', ``operating cash flow'', ``cash'', ``share repurchases'', and ``dividends'' -- metrics that also appear among the most frequent targets in Tab.~\ref{tab:freq}.
This example demonstrates that the LLM's ability to capture contextually meaningful metrics at the individual level translates to more informative signals at the aggregate level.

\paragraph{LLM-based method yields stronger portfolio returns than the NER-based baseline.}
Tab.~\ref{tab:portfolio} presents calendar-time portfolio excess returns and alphas sorted by the moving targets score.
Panel (A) shows that the NER-based method produces a negative Q5--Q1 spread: excess return of $-0.31\%$ ($t=-1.48$), 3-factor alpha of $-0.18\%$ ($t=-0.89$), and 5-factor alpha of $-0.14\%$ ($t=-0.70$).
In contrast, Panel (B) shows that the LLM-based method yields a significantly negative spread: excess return of $-0.41\%$ ($t=-2.08$), 3-factor alpha of $-0.48\%$ ($t=-2.40$), and 5-factor alpha of $-0.52\%$ ($t=-2.55$), all significant at the 5\% level.
This indicates that firms with high target shifting reliably underperform those with low target shifting.
The stronger predictability suggests that LLM-based extraction captures semantically meaningful changes that the NER-based approach misses.

\begin{table}[t]
\centering
\caption{
Fama–MacBeth regressions of individual firm-level stock returns on \emph{Moving Targets} and standard return predictors.
(A) reports results using the NER-based extraction method, while (B) reports results using the LLM-based extraction method.
We include standard controls for known return predictors: Log(Size), Log(BM), Ret(-1,0), and Ret(-12,-1).
$t$-statistics are shown in parentheses.
}
\label{tab:regression}
\resizebox{1.0\columnwidth}{!}{
\begin{tabular}{lcc}
\toprule
 & (A) & (B) \\
 & \textcolor{ner}{\emph{NER-based method}} & \textcolor{llm}{\emph{LLM-based method}} \\
\midrule
 & \multicolumn{2}{c}{Ret} \\
\midrule
Mt Score    & 0.0107 & \textbf{-0.0370} \\
            & (1.10) & (-0.95) \\
Log (Size)    & -0.0015 & -0.0072 \\
            & (-0.42) & (-1.37) \\
Log (BM)      & -0.0016 & -0.0037 \\
            & (-0.36) & (-0.71) \\
Ret (-1,0)    & 0.0404 & -0.0501 \\
            & (0.98) & (-0.81) \\
Ret (-12,-1) & -0.0019 & 0.0006 \\
            & (-0.22) & (0.06) \\
Constant    & 0.0129 & 0.1200 \\
            & (0.29) & (1.53) \\
\midrule
R-squared   & 0.3897 & 0.3915 \\
N           & 5{,}615 & 5{,}615 \\
\bottomrule
\bottomrule
\end{tabular}
}
\end{table}

\paragraph{LLM-based method shows more predictive power in cross-sectional regressions.}
Tab.~\ref{tab:regression} reports Fama-MacBeth regression results testing whether the moving targets score predicts next-month stock returns.
Column (A) shows that the NER-based method rather yields a positive coefficient on the moving targets score, suggesting no reliable predictive relationship.
In contrast, Column (B) shows that the LLM-based method produces a negative coefficient of $-0.0370$, indicating that higher target shifting predicts lower subsequent returns even after controlling for known return predictors.
These results are consistent with the portfolio analysis and further confirm that LLM-based extraction captures return-relevant information that the NER-based approach does not.

\section{Conclusion}
We investigated whether LLMs can effectively track evolving signals in corporate disclosures by comparing an LLM-based framework against a NER-based baseline on the moving targets phenomenon.
Our results show that LLM-based extraction combined with semantic similarity yields stronger return predictability: the long-short portfolio achieves a significant five-factor alpha of $-0.52\%$, and Fama-MacBeth regressions confirm significant predictive power at the 10\% level.
In contrast, the NER-based baseline produces insignificant results in both analyses.
Qualitative examination reveals two sources of improvement.
First, LLMs preserve contextual qualifiers that distinguish semantically related metrics (e.g., ``Total revenue'' vs.\ ``Blackwell revenue''), which NER collapses into generic labels.
Second, LLMs avoid extracting non-metric terms that introduce noise into the moving targets calculation.
Our findings support the view that LLMs can capture meaning beyond surface-level wording, offering a more robust approach to quantifying semantic changes in financial text.
While we focus on earnings call transcripts, the framework is portable to other corporate disclosures rich in evolving signals, such as annual reports and shareholder letters.
\newpage
\section*{Limitations}

We restrict our sample to S\&P 100 firms due to computational costs. Extending the analysis to a broader set of firms, particularly smaller firms with different disclosure practices, would be a valuable direction for future work.
In addition, our framework depends on the quality of LLM extraction and embedding models; evaluating across different model choices would help verify robustness.
Lastly, we demonstrate predictive gains but do not investigate why LLM-based tracking better captures return-relevant information; understanding the underlying economic mechanisms is a valuable direction for future work.


\bibliography{main}

\appendix
\onecolumn
\section{Appendix}
\label{sec:appendix}

\begin{table*}[t]
\centering
\caption{
LLM prompt templates used in our method to extraction of targets.
Note that the terms \emph{label} and \emph{target} are used in the prompt to refer to what we denote as \emph{metrics} in the main text.
}
\label{tab:prompts}
\begin{tabular}{p{0.1\textwidth} p{0.9\textwidth}}
\toprule
\textbf{System Prompt} & 
\begin{minipage}[t]{\linewidth}\small
\lstset{breaklines=true, breakatwhitespace=true, columns=fullflexible}
\begin{lstlisting}
<task>
From the provided inputs, extract all performance metric and target labels
and classify them into two separate lists according to the section
in which they were mentioned:
- presentation: items mentioned in the prepared presentation or opening remarks
- analyst_qa: items mentioned during analyst question and answer sessions
The term "performance metric and target labels" includes every explicitly stated
financial result, non-financial result, and key business indicator discussed or mentioned. Both GAAP and non-GAAP labels if explicitly distinguished.
## Section classification
- presentation: items mentioned in prepared remarks/opening statements before Q&A begins.
  Typical speakers: executives with "- Executives" in the name, before the first analyst question.
- analyst_qa: items mentioned during the Q&A (analyst questions and executive answers after Q&A starts).
  Q&A typically starts after an Operator prompt like "[Operator Instructions]" or when an analyst first speaks
  (speaker label contains "- Analysts") and continues until the call ends.
- Operator-only lines do not themselves contain items, but they can mark Q&A start.
- Include only labels explicitly stated in content
- Deduplicate near-duplicates to one normalized target per section
</task>
<inputs>earnings-call transcript as indexed JSON dialog</inputs>
<output>
- target:
  - Short noun phrase only
  - **Quarter results:** prefix with "Quarterly" and do NOT include any years or quarter numbers
  - **Fiscal-year results:** prefix with "Yearly" and do NOT include any years
  - **Guidance/targets/other indicators:** no period prefixes unless inherently part of the label; avoid dates, years, and quarter numbers
  - No numbers, units, currency symbols, or percent signs anywhere
  - Prefer consistent terminology across products/segments
- index: integer index of the utterance containing the quote
</output>
<format>
{{
    "presentation": List[Dict],
    "analyst_qa": List[Dict]
}}
The dictionary must follow the following structure:
{{
    "target": str # Normalized short label for the metric/target according to the rules
    "index": int # Zero-based index of the utterance in the transcript where the quote appears
}}
</format>
\end{lstlisting}
\end{minipage} \\
\bottomrule
\end{tabular}
\end{table*}
Table~\ref{tab:prompts} shows a full system prompt used to extract metrics in our proposed method.

\end{document}